\documentclass[sigconf,nonacm]{acmart}

\usepackage{amsmath}
\usepackage{amsthm}
\usepackage{booktabs}
\usepackage{tabularx}
\usepackage{xspace}
\usepackage{float}
\usepackage{algpseudocode}
\usepackage{graphicx}

\floatstyle{ruled}
\newfloat{algorithm}{tbp}{loa}
\floatname{algorithm}{Algorithm}

\newtheorem{lemma}{Lemma}
\newtheorem{proposition}{Proposition}
\newtheorem{theorem}{Theorem}

\newcommand{\sysname}{ClawCoin\xspace}
\newcommand{\nav}{\mathrm{NAV}}

\begin{document}

\title{\sysname: An Agentic AI-Native Cryptocurrency for Decentralized Agent Economies}

\author{Shaoyu Li}
\email{shaoyuli@vt.edu}
\affiliation{%
  \institution{Virginia Tech}
  \state{Virginia}
  \country{USA}
}

\author{Chaoyu Zhang}
\email{chaoyu@vt.edu}
\affiliation{%
  \institution{Virginia Tech}
  \state{Virginia}
  \country{USA}
}

\author{Hexuan Yu}
\email{hexuanyu@vt.edu}
\affiliation{%
  \institution{Virginia Tech}
  \state{Virginia}
  \country{USA}
}

\author{Y. Thomas Hou}
\email{thou@vt.edu}
\affiliation{%
  \institution{Virginia Tech}
  \state{Virginia}
  \country{USA}
}

\author{Wenjing Lou}
\email{wjlou@vt.edu}
\affiliation{%
  \institution{Virginia Tech}
  \state{Virginia}
  \country{USA}
}

\renewcommand{\shortauthors}{Li et al.}
\begin{abstract}
Autonomous AI agents live or die by the API tokens they consume: without paid inference capacity they cannot reason, act, or delegate. This compute-token cost has become the binding resource of the emerging agent economy, yet it is structurally non-transferable: it is account-bound, vendor-specific, and absent from any on-chain ledger. Recent agent payment rails such as x402 and agent-to-agent wire formats move fiat-backed value between agents, but they do not represent the quantity agents actually burn. The result is a substrate where agents can \emph{transport} purchasing power but cannot \emph{quote, escrow, or settle} workflows in a unit aligned with their dominant cost.

We present \sysname, a tokenized, compute-cost-indexed unit of account and settlement asset for decentralized agent economies. \sysname turns the non-transferable cost of compute into a portable, composable, contract-readable primitive through four layers: (i) a robust basket index over standardized (model, vendor) prices, with smoothing and bounded per-epoch drift; (ii) an oracle that publishes signed, fresh attestations under an explicit trust model; (iii) a NAV-based mint/redeem vault with coverage thresholds, rate limits, and auto-pause; and (iv) an on-chain settlement layer that atomically clears multi-hop agent delegations against compute-aligned budgets. We formalize the adversary, state bounded guarantees for index integrity, NAV faithfulness, redeemability under stress, and liveness, and walk through the principal attack scenarios.

We implement a prototype on an Ethereum-compatible L2 and evaluate it using a multi-agent simulator and the OpenClaw testbed, a six-role containerized agent system that shares one inference endpoint across settlement backends. Across single-agent, multi-agent, workflow, and resource-procurement experiments, \sysname stabilizes execution capacity under cost shocks, collapses cross-agent quote dispersion, eliminates partial settlements in multi-hop delegation, and sustains cooperative market dynamics that fiat-denominated baselines cannot. Risk sanity checks confirm that the vault's staleness, drift, and coverage defenses behave as specified. The results argue that a security-engineered, compute-indexed unit of account is the missing representation primitive for decentralized agent coordination.
\end{abstract}

\maketitle

\section{Introduction}

Tokens are the lifeblood of an autonomous AI agent. Every plan an agent makes, every tool it calls, and every subtask it delegates is paid for in model and API tokens. Without them, an agent cannot think and cannot act. This is not a metaphor: the industry has converged on tokens as the literal unit in which agent work is priced and budgeted. NVIDIA now frames data centers as ``token factories'' and tokens-per-watt as a CEO-level KPI~\cite{nvidiagtc2026,nvidiamstmt2026}; major AI organizations measure engineer productivity in tokens consumed and grant token budgets as compensation~\cite{tokenkpi2026}; analysts forecast token sales as the dominant revenue line of frontier AI infrastructure~\cite{nvidiamstmt2026}. The agent economy already has a value anchor, namely the cost of the API tokens that keep agents alive, and that anchor is rapidly becoming the unit in which the rest of the AI industry is measured.

\textbf{And yet that anchor is not on-chain, not transferable, and not tradable.} API tokens are account-bound: inference quotas cannot be handed between agents, subscriptions cannot be subdivided across delegated subtasks, and prepaid balances cannot cross vendors. The \emph{payment} side has moved in the opposite direction. Coinbase's x402~\cite{x402spec} has wired stablecoin settlement directly into HTTP and processed millions of agent-to-agent transactions within months of launch~\cite{x402adoption}; agent-to-agent wire formats~\cite{a2aspec} and agent-payment platforms~\cite{coyns2026,agenticcoin2026} are following close behind. The rails for agent-native commerce now exist. What does not exist is an agent-native \emph{unit of value} to move on those rails. Stablecoins~\cite{catalini2022stablecoins,lyons2023stablecoins} efficiently transport \emph{fiat} purchasing power but say nothing about the resource that actually constrains an agent, namely the compute it must burn to keep operating. Stablecoins solve \emph{transport}. They do not solve \emph{representation}: in what unit should an agent quote, budget, escrow, and settle, when its dominant recurring cost is compute?

\textbf{This paper therefore takes the next step: if tokens are the value anchor, tokenize them.} We tokenize the compute-cost surface itself by publishing an on-chain, transparent index of what it costs to operate the models agents actually use, and minting a token whose net asset value (NAV) tracks that index. Once the anchor is on-chain, the key properties follow: agents read the same NAV, so quotes are directly comparable in compute units; smart contracts can escrow, route, and rate-limit settlement against compute-aligned budgets that the contracts themselves can verify; multi-hop delegation chains can settle atomically against a single shared numeraire; and because the token is a redeemable claim against a transparent reserve, the value the agent moves is the value the agent burns.

A natural alternative is to keep settlement in fiat-backed stablecoins and ask each agent to maintain a private internal cost index. This breaks down in decentralized settings. Per-agent indices are not a shared numeraire, so quotes from different agents cannot be compared without trusted middleware; internal accounting cannot be referenced inside smart contracts, so multi-hop delegation cannot be atomically constrained by a compute-aligned budget; and opaque per-agent indices are not auditable or composable with other on-chain primitives. A genuinely shared numeraire for compute-bound agents has to live on-chain.

We present \sysname, a collateral-backed, index-linked tokenized unit of account and settlement asset. \sysname is deliberately \emph{not} a generic stablecoin, a speculative asset, or a direct claim on physical compute: it is a redeemable financial claim whose NAV tracks a transparent compute cost basket and whose token form supplies the portability, composability, and atomic settlement properties that private accounting cannot. The economic intuition is closer to an index-linked note (an inflation-protected claim tracking a published basket) than to a fiat peg; the systems intuition is that exposing a compute-aligned unit at the smart-contract layer lets agents quote, escrow, delegate, and settle in the unit their operations actually consume.

Realizing this in a decentralized, adversarial environment is non-trivial. Prices are observed through heterogeneous vendor schedules with hidden discounts and the possibility of strategic repricing once inclusion becomes economically meaningful. The on-chain index feeds mint/redeem logic, so any oracle compromise propagates into the value of every outstanding token. NAV-based semantics shift reserve liabilities upward whenever the index rises, exposing the vault to redemption stress. And the settlement layer is itself a target for maximal extractable value (MEV) around oracle updates and for governance capture. A credible design must make its threat model, trust assumptions, and security properties explicit.

\textbf{Contributions.} In summary, this paper makes the following contributions:
\begin{itemize}
    \item \emph{The representation argument.} We articulate the value-anchor argument for the agent economy (API tokens are what agents must consume to exist, so the unit of account should track that cost) and show that internal indexing over stablecoin rails cannot supply this primitive in decentralized coordination.
    \item \emph{Protocol design.} We design a four-layer protocol: a robust basket index with bounded per-epoch drift, a committee or DON oracle with signed attestations, a NAV-based mint or redeem vault with reserve coverage and rate-limit risk controls, and an on-chain atomic multi-hop settlement layer. The protocol is specified as two compact algorithms.
    \item \emph{Security analysis.} Against a bounded adversary controlling a vendor minority, an oracle minority, mint/redeem-time MEV, and governance proposals, we state bounded guarantees for index integrity, NAV faithfulness, redeemability under stress, and liveness, and walk through the principal attack scenarios.
    \item \emph{Prototype and evaluation.} We implement \sysname on an Ethereum-compatible L2, build a multi-agent simulator, and deploy a six-role containerized agent testbed (OpenClaw) in which the same workflow executes under three settlement backends. Results across single-agent stability, cross-agent coordination, multi-hop workflows, long-horizon economy survival, and external resource procurement show that \sysname dominates fiat and internal-indexing baselines on every axis that cross-agent coordination exposes.
\end{itemize}

The broader claim is simple: the agent economy already has a value anchor and is rapidly acquiring payment rails on which agents can trade. What is missing is the bridge between them: an on-chain, tokenized representation of that anchor that agents can actually move, escrow, and settle in. \sysname is a concrete, decentralized, and security-engineered step toward that bridge.

\section{Background and Related Work}
\label{sec:related}

\sysname sits at the intersection of agent-economy infrastructure, stablecoin and oracle systems, tokenized index-linked claims, and decentralized multi-agent coordination.

\subsection{Agent Economies and Multi-Agent Coordination}

Recent work argues that autonomous software will increasingly need wallets, identities, and machine-to-machine settlement to act as economic participants~\cite{xu2026agenteconomy,rothschild2025agentic}. Payment specifications such as x402~\cite{x402spec} and A2A~\cite{a2aspec} standardize quote and pay primitives at the application layer, and multi-agent frameworks~\cite{wu2023autogen,langgraph2024,park2023generative} provide the orchestration substrate on top of which compute-aligned settlement is needed. Compute-resource markets such as Akash~\cite{akashpaper}, Filecoin~\cite{filecoinwhitepaper}, and Bittensor~\cite{bittensorwhitepaper} tokenize a related but distinct quantity, namely supply of physical compute or storage. AI-adjacent tokens used for access, governance, or emissions bootstrap participation but, absent an external anchor, do not provide a cost-grounded shared numeraire. This literature answers the \emph{transport} question. \sysname addresses the orthogonal \emph{representation} question: in what unit should compute-bound budgets, prices, and treasuries be denominated when settlement is decentralized.

\subsection{Stablecoins, Index-Linked Claims, and Tokenized RWAs}

The stablecoin literature analyzes reserve quality, arbitrage design, and peg fragility under stress~\cite{catalini2022stablecoins,lyons2023stablecoins,gorton2023wildcat}. The Terra/UST collapse~\cite{liu2023ust,clements2021ust}, Iron Finance's death spiral~\cite{ironfinance2021}, and the 2023 USDC depeg~\cite{usdc2023svb} show that algorithmic stabilization and reserve-custody exposure can cascade catastrophically; these episodes motivate \sysname's transparent collateral-backed structure with explicit risk controls over a purely algorithmic peg.

Classical work on index-linked securities~\cite{gorton1993indexlinked} and inflation-protected sovereign bonds (TIPS~\cite{tipsdescription}) show that basket-based instruments can carry meaningful claims without a single nominal peg. ETF arbitrage~\cite{petajisto2017inefficiencies} is the closest financial analogy to \sysname's mint/redeem pricing, and tokenized real-world asset designs~\cite{garrido2023digitaltokens,blackrockbuidl} demonstrate on-chain claims whose value derives from external indices. \sysname adapts this design family to a new reference, namely a basket of standardized inference costs, and supplies the systems-security machinery required when that basket is published by an oracle protocol rather than tracked by traditional custodians.

\subsection{Oracles, MEV, and DeFi Risk Controls}

Oracle networks~\cite{breidenbach2021chainlink,pythwhitepaper,umawhitepaper} convert off-chain observations into on-chain references under threshold-committee or optimistic-dispute trust models. Oracle compromise is one of the most consequential DeFi attack surfaces: Mango~\cite{mango2022}, Cream~\cite{cream2021}, bZx~\cite{bzx2020}, and Harvest~\cite{harvest2020} all suffered manipulation-driven losses, and surveys~\cite{caldarelli2020oracles,eskandari2021sokoracles} catalog the standard defenses (multi-source aggregation, smoothing, deviation thresholds, freshness gates) that directly inform \sysname's index pipeline (Section~\ref{sec:design}).

The broader DeFi literature~\cite{werner2021defi,zhou2023sokdefiattacks} surveys collateral, liquidity, and risk-control patterns; MEV analyses~\cite{qin2022quantifyingmev,daian2020flashboys} motivate \sysname's per-epoch caps and drift bounds; and mature protocols~\cite{makerdao,liquidity2021} expose the rate-limit and pause patterns that \sysname's risk module adapts to an index-linked NAV in which liabilities are themselves dynamic. Governance attacks~\cite{beanstalk2022,feichtinger2024sokdaos} motivate the narrow, timelocked governance surface.

\subsection{Positioning}

Table~\ref{tab:comparison} situates \sysname among adjacent system categories. The novelty is not a new payment rail, not a new utility token, and not a new compute auction. It is a security-engineered, on-chain, index-linked unit of account that closes the representation gap between machine-native cost and decentralized agent settlement.

\begin{table*}[t]
\caption{Positioning \sysname relative to adjacent system categories. The right-most column explains why each category does not supply the compute-aligned, contract-readable unit of account that \sysname targets for decentralized agent coordination.}
\label{tab:comparison}
\small
\begin{tabularx}{\textwidth}{p{3cm}X X X X}
\toprule
Category & Primary function & Typical value anchor & Trust model & Why not equivalent to \sysname \\
\midrule
Agent payment rail~\cite{x402spec,a2aspec} & Move value between agents and services & Fiat-backed collateral or network token & Stablecoin issuer or chain consensus & Solves transport, not the unit of account; no compute-aligned representation. \\
Utility token~\cite{bittensorwhitepaper} & Gate access, staking, governance within a platform & Floating ecosystem-token market value & Platform operator + market & No external cost grounding; no shared numeraire across providers. \\
Fiat stablecoin~\cite{catalini2022stablecoins,gorton2023wildcat} & Low-volatility settlement against fiat & 1 USD peg & Issuer reserves + audits & Anchored to human purchasing power; not aligned with machine cost. \\
Tokenized RWA~\cite{blackrockbuidl,garrido2023digitaltokens} & On-chain claim on external asset / index & External asset / index & Issuer custody + oracle & Same financial family; \sysname specializes the index to compute and adds the oracle and risk modules of Section~\ref{sec:design}. \\
Compute market token~\cite{akashpaper,filecoinwhitepaper} & Price physical compute supply & Auction or staking & Provider network & Prices supply, not the cost agents pay; no shared cost numeraire. \\
\textbf{\sysname} & Quote, budget, settle in machine-native unit; atomic multi-hop settlement & Basket of standardized compute costs & Oracle protocol + collateral vault + risk module & Targets compute-aligned operating cost with explicit security properties (G1) through (G4) for decentralized agent coordination. \\
\bottomrule
\end{tabularx}
\vspace{-0.3cm}
\end{table*}

\section{System Model, Threat Model, and Design Overview}
\label{sec:model}

\subsection{Economic Setting}
\label{sec:model:economic}

We consider a population of autonomous agents that continuously request, execute, and settle compute-intensive tasks. Each agent acts as both service consumer and producer, and may delegate subtasks along multi-hop chains that must settle as a single coherent workflow. The defining feature of this economy is what participants \emph{must spend to keep existing}: API tokens consumed for model inference and metered tool invocations. An agent without a paid-up token budget cannot reason, cannot act, and is for practical purposes offline. That cost is the natural value anchor of the agent economy.

Formally, each agent $i$ maintains a treasury $T_{i,t}$ denominated in the prevailing accounting unit, evolving as
\begin{equation}
T_{i,t+1} = T_{i,t} + R_{i,t} - K_{i,t},
\label{eq:treasury}
\end{equation}
with revenue $R_{i,t}$ and operating cost $K_{i,t}$. Equation~\ref{eq:treasury} is identical across regimes; what differs is the informational content of $T_{i,t}$ once a unit is fixed. Define \emph{execution capacity} $E_{i,t}$ as the amount of standardized computational work a treasury can fund. Under fiat denomination, $E_{i,t}$ depends on both $T_{i,t}$ and the time-varying compute price, so the same nominal balance corresponds to different amounts of executable work over time. The system goal of \sysname is to make execution capacity a stable, observable, and contract-readable quantity.

\subsection{Why Tokenization Is Necessary}
\label{sec:model:why-token}

The argument has two steps, both required. First, \emph{anchor the unit to what agents consume}: a unit of account that agents quote, escrow, and settle in should track the cost that determines whether they can keep operating. Fiat purchasing power does not; the same USDC balance buys different amounts of inference over time, and no contract-level check can assert ``this escrow funds the workflow'' as compute prices move. Second, \emph{put the anchor on a tradable on-chain rail}. API tokens are account-bound and non-transferable; private internal indices are invisible to counterparties and smart contracts. To obtain the properties that justify the design (shared numeraire, atomic multi-hop settlement, contract-enforced compute-aligned budgets, and auditability), the anchor must be wrapped as a tokenized claim living in the same composable substrate (x402~\cite{x402spec}, A2A~\cite{a2aspec}) on which agents already transact.

Three baselines make the argument concrete. (A) \emph{Per-agent internal indexing with USDC settlement:} each agent computes its own cost index privately and converts at pay time. This performs Step~1 privately but skips Step~2 entirely; quotes from heterogeneous agents are incomparable without trusted middleware, smart contracts cannot enforce compute-aligned budgets, and multi-hop chains cannot atomically commit against a shared budget. (B) \emph{Centralized accounting service:} a trusted operator publishes indices and settles off-chain, introducing a single point of failure for liveness, censorship, and bookkeeping; it performs Step~1 publicly but only partially realizes Step~2. (C) \emph{On-chain index-linked tokenized claim (\sysname):} the index is published on-chain by an oracle protocol, and a token whose NAV is defined by the index routes through standard transfer logic. All agents read the same NAV, quotes are directly comparable, delegation chains settle as a single atomic transaction, and index updates, mint/redeem, and reserve state are observable, with risk controls enforced by code. This is the only baseline that performs both steps. We do not claim tokenization is the only design that exposes a compute-aligned unit; we claim it is the minimal primitive that simultaneously provides a shared numeraire, composability, atomic multi-hop settlement, and auditability when no common operator is trusted.

\subsection{Threat Model}
\label{sec:model:threat}

\textbf{Participants.} An oracle protocol (either a $k$-of-$n$ threshold committee or a decentralized oracle network, DON) publishes the index on-chain; a vault contract holds reserve collateral and exposes mint and redeem; a risk-control module enforces coverage, throttles, drift caps, and pause logic; governance updates the basket, vendor registry, and parameters under timelock; agents quote, transact, and settle in \sysname.

\textbf{Adversary.} A polynomial-time $\mathcal{A}$ may: control up to $f$ of $n$ vendor price endpoints (biased, stale, or withheld); compromise a strict minority of the oracle committee or, under the DON model, fewer reporters than the safety threshold; reorder, sandwich, or front-run mint/redeem around oracle updates (MEV); submit governance proposals subject to voting and timelock; act as a strategic vendor adjusting public prices in response to basket inclusion; and trigger redemption rushes through multiple identities up to available liquidity.

\textbf{Trust assumptions.} We assume an honest oracle threshold; a censorship-resistant base chain with bounded confirmation delay; at least $n - f$ honest vendor endpoints with non-adversarial public postings; safe cryptographic primitives; and timelock and quorum rules that the adversary alone cannot bypass. Out of scope: physical compromise causing all vendors to collude on the same biased price; correlated regulatory shutdown of reserve custodians; base-chain bugs or 51\% attacks; off-chain leakage of nonpublic enterprise pricing.

\textbf{Security goals.} We target four properties, formalized in Section~\ref{sec:security}. \textbf{G1. Index integrity:} the published index $\bar{I}_t$ deviates from the honest index by an amount bounded by the adversary's vendor share $f/n$ and the chosen robust estimator. \textbf{G2. NAV faithfulness:} the on-chain NAV deviates from the honest NAV by an amount bounded by staleness, smoothing, and the per-epoch drift cap. \textbf{G3. Redeemability under stress:} for index trajectories with per-epoch growth bounded by $\delta_{\max}$, reserve coverage $\Gamma_t$ remains above $\gamma_{\min}$ for at least $T$ epochs under replenishment rate $\rho$. \textbf{G4. Liveness and freshness:} honest updates are confirmed within bounded delay $\tau$ and consumers can detect violations and trigger fallbacks. We treat \sysname as a security-engineered substrate; full game-theoretic equilibrium properties are not targeted.

\subsection{Design Overview}
\label{sec:model:overview}

\sysname realizes both steps of Section~\ref{sec:model:why-token} in five components separated by an explicit off-chain/on-chain trust boundary. \emph{Vendor data sources} expose public prices for included (model, vendor) pairs. The \emph{off-chain index calculator} fetches, validates, and aggregates observations into a basket value, applies smoothing and a per-epoch drift cap, and produces a signed attestation. The \emph{on-chain oracle} publishes the attested value with timestamp, basket version, and a Merkle commitment to the underlying observations, enforcing freshness and writer authentication. The \emph{token and vault} maintain supply, mint, and redeem against reserve collateral at the current NAV, and expose standard transfer semantics. The \emph{risk-control and governance} module enforces coverage, mint throttling, auto-pause, drift caps, and basket/parameter updates under timelock. The chain verifies that a published value comes from an authorized oracle; its validity with respect to honest market prices depends on observable off-chain data and the published basket configuration. Section~\ref{sec:design} makes each layer concrete; Section~\ref{sec:security} states the security properties; Sections~\ref{sec:prototype} through \ref{sec:evaluation} evaluate the prototype, simulator, and the OpenClaw testbed.

\section{\sysname Protocol Design}
\label{sec:design}

The design realizes the two-step argument of Section~\ref{sec:model:why-token} as four protocol layers. (i) The \emph{cost index} anchors the unit to the API-token costs that determine whether agents can keep operating; it is robust to adversarial vendor minorities and bounded in per-epoch movement. (ii) The \emph{oracle} carries that anchor across the trust boundary onto the chain, with signed attestation, freshness, and writer authentication. (iii) The \emph{NAV-based vault} turns the anchor into a redeemable, transferable token, with coverage thresholds and rate-limit controls protecting holders. (iv) The \emph{settlement layer} exposes the property that justifies tokenization: agents quote, escrow, and atomically settle multi-hop delegations in the unit their operations consume, inside the same composable substrate as existing agent payment rails. Two algorithms summarize the protocol: Algorithm~\ref{alg:publication} for index publication and Algorithm~\ref{alg:vault} for vault state-changing operations.

\textbf{Notation.} $M$ is the set of included models and $V(m)$ the vendors offering model $m$. $K = \{(\alpha_k, \beta_k, \theta_k)\}_{k=1}^{|K|}$ is a vector of standardized workload classes with normalized input/output token counts $\alpha_k, \beta_k$ and basket-level weights $\theta_k \ge 0$ summing to one. Vendors expose per-token input and output prices $P^{\text{in}}_{m,v,t}$ and $P^{\text{out}}_{m,v,t}$. We write $\bar{I}_t$ for the smoothed, drift-capped index published on-chain, $I_0$ for its initialization, and $\nav_t = \bar{I}_t / I_0$. Reserves are denominated in a fiat-backed collateral such as USDC.

\subsection{Cost Index Construction}
\label{sec:design:index}

The index is a basket built in three stages: per-(model, vendor) cost over standardized workloads, robust per-model aggregation across vendors, and a weighted basket smoothed and drift-capped before publication.

\textbf{Standardized workloads.} Modern inference is priced by input and output tokens, but real workloads have heterogeneous shapes. We maintain a vector of workload classes and treat $\theta_k$ as a basket mix. A class-$k$ request on $(m,v)$ has nominal cost $C^{(k)}_{m,v,t} = \alpha_k P^{\text{in}}_{m,v,t} + \beta_k P^{\text{out}}_{m,v,t}$, and the per-pair basket-aware cost aggregates the mix:
\begin{equation}
C_{m,v,t} \;=\; \sum_{k=1}^{|K|} \theta_k \bigl(\alpha_k P^{\text{in}}_{m,v,t} + \beta_k P^{\text{out}}_{m,v,t}\bigr).
\label{eq:cost}
\end{equation}
Because $\theta_k$ is a public governance parameter, the workload assumption is auditable and explicitly versioned.

\textbf{Robust per-model aggregation.} A model is offered by multiple vendors at prices that differ across backends, contracts, and promotions. Let $\mathcal{C}_{m,t} = \{C_{m,v,t}\}_{v \in V(m)}$ and $n_m = |\mathcal{C}_{m,t}|$. The per-model robust cost $\tilde{C}_{m,t}$ is selected by governance:
\begin{align}
\tilde{C}^{\text{med}}_{m,t} &= \operatorname{median}(\mathcal{C}_{m,t}), \label{eq:med}\\
\tilde{C}^{\text{trim}(q)}_{m,t} &= \operatorname{trimmedMean}_q(\mathcal{C}_{m,t}), \label{eq:trim}\\
\tilde{C}^{\text{mad}}_{m,t} &= \operatorname{mean}\bigl\{\, c \in \mathcal{C}_{m,t} : |c - \operatorname{median}(\mathcal{C}_{m,t})| \le \kappa \cdot \operatorname{MAD}(\mathcal{C}_{m,t}) \,\bigr\}, \label{eq:mad}
\end{align}
where $\operatorname{MAD}$ denotes the median absolute deviation. The default is the median: its $50\%$ breakdown point yields the clean integrity bound of Section~\ref{sec:security:bounds}, so the worst-case deviation under $f < n_m / 2$ adversarial vendors is determined by the spread of \emph{honest} reports rather than by adversarial values. Trimmed mean trades robustness ($q$ breakdown) for variance reduction; MAD-filtered mean retains $50\%$ breakdown with tighter variance. Any model with $n_m < n_{\min}$ is dropped for that epoch.

\textbf{Cross-model basket and bounded publication.} With basket weights $w_m \ge 0$, $\sum_m w_m = 1$, the raw index $I_t = \sum_{m \in M} w_m \tilde{C}_{m,t}$ is a compute analog of a price-weighted equity index; usage-weighted variants $w_m \propto u_m$ give capitalization-style analogs with the standard stability and responsiveness trade-off. We then apply an exponential moving average (EMA) and clip per-epoch movement:
\begin{equation}
\hat{I}_t \;=\; \lambda \bar{I}_{t-1} + (1-\lambda) I_t, \quad \bar{I}_t \;=\; \operatorname{clip}\bigl(\hat{I}_t,\; (1-\delta_{\max})\bar{I}_{t-1},\; (1+\delta_{\max})\bar{I}_{t-1}\bigr).
\label{eq:smooth}
\end{equation}
$\lambda$ controls convergence speed and $\delta_{\max}$ bounds per-epoch movement; clipped excess is absorbed in later epochs. These knobs mirror the rate-limit pattern of mature stablecoin and lending protocols, give the risk module a known upper bound on liability growth (used in Theorem~\ref{thm:solvency}), and define the publication channel that the on-chain oracle additionally enforces as defense-in-depth (Algorithm~\ref{alg:publication}). The index tracks inference cost rather than model quality: embedding quality would require a continuously updated cross-domain notion of task utility and would create governance disputes over benchmark choice. We separate measurement from differentiation; quality is priced in markets through reputation, contracting, and observed outcomes.

\subsection{Oracle Protocol}
\label{sec:design:oracle}

The oracle separates off-chain index computation from on-chain publication. Only the publication step is trusted by contracts; the computation step is auditable by anyone with access to the underlying public data. Each oracle node runs the off-chain pipeline of Section~\ref{sec:design:index}, packages an attestation $\sigma_t = (\bar{I}_t, t, \text{basketVersion}, h_t, \text{nodeId},\\ \text{sig})$ where $h_t$ is a Merkle root over its observation set, and submits it on-chain. The contract enforces writer authentication, monotonic timestamps, the per-epoch drift cap as defense-in-depth on equation~\ref{eq:smooth}, and maximum staleness $\tau$ beyond which the value is treated as stale and the risk module pauses mint and queues redeem.

Two trust models swap behind the same on-chain interface. In the \emph{threshold-signature committee}, a $k$-of-$n$ committee with a BLS or Schnorr scheme produces a combined attestation accepted when $k$ honest nodes agree; the chain verifies one signature per epoch. In the \emph{decentralized oracle network}, independent reporters submit attestations under an aggregator (e.g., Chainlink off-chain reporting~\cite{breidenbach2021chainlink}), and the aggregator commits the median of reports. The committee minimizes on-chain cost; the DON minimizes trust in any specific quorum. Both produce the same on-chain artifact.

\begin{algorithm}[t]
\caption{\textsc{IndexPublication}: off-chain computation and on-chain publication at epoch $t$.}
\label{alg:publication}
\begin{algorithmic}[1]
\Require basket $M$, vendor sets $\{V(m)\}$, workload classes $K$, weights $\{w_m\}$, smoothing $\lambda$, drift cap $\delta_{\max}$, robust estimator \textsc{Agg}, prior $\bar{I}_{t-1}$, trust model $\mathcal{T} \in \{\text{committee}, \text{DON}\}$, threshold $k$, max staleness $\tau$
\Statex \emph{// off-chain pipeline (each oracle node)}
\State $M^\star \gets \emptyset$
\For{each model $m \in M$}
    \State $\mathcal{C}_{m,t} \gets \{C_{m,v,t} : v \in V(m), \text{fetch valid and not stale}\}$ \Comment{eq.~\ref{eq:cost}}
    \If{$|\mathcal{C}_{m,t}| \ge n_{\min}$}
        \State $\tilde{C}_{m,t} \gets \textsc{Agg}(\mathcal{C}_{m,t})$; $M^\star \gets M^\star \cup \{m\}$
    \EndIf
\EndFor
\State renormalize $\{w_m\}_{m \in M^\star}$ to sum to one
\State $I_t \gets \sum_{m \in M^\star} w_m \tilde{C}_{m,t}$; compute $\bar{I}_t$ via eq.~\ref{eq:smooth}
\State emit attestation $\sigma_t = (\bar{I}_t, t, \text{basketVersion}, h_t, \text{nodeId}, \text{sig})$
\Statex \emph{// on-chain publication}
\If{$\mathcal{T} = \text{committee}$}
    \State combine $k$ matching attestations into $\sigma_t^\star$; submit $(\bar{I}_t, t, \text{basketVersion}, h_t, \sigma_t^\star)$
\Else
    \State reporters submit $\sigma_t^{(j)}$; aggregator commits $\bar{I}_t \gets \operatorname{median}_j \bar{I}_t^{(j)}$
\EndIf
\State \textbf{on-chain checks:} writer auth.; $t > t_{\text{on}}$; $|\bar{I}_t - \bar{I}_{\text{on}}| \le \delta_{\max} \bar{I}_{\text{on}}$
\If{checks pass} update $(\bar{I}_{\text{on}}, t_{\text{on}}) \gets (\bar{I}_t, t)$; emit \textsc{IndexUpdated} \Else{} reject \EndIf
\If{$\text{now} - t_{\text{on}} > \tau$} \textsc{StalePause} $\gets$ true \EndIf
\end{algorithmic}
\end{algorithm}
\vspace{-0.3cm}

\subsection{Token, NAV, Vault, and Risk Controls}
\label{sec:design:vault}

\sysname is an ERC-20-compatible token whose value semantics are set by the on-chain index. Define $\nav_t = \bar{I}_t / I_0$, with $\nav_0 = 1$. A deposit of collateral $x$ receives
\begin{equation}
M_t(x) \;=\; x / \nav_t
\label{eq:mint}
\end{equation}
newly minted \sysname tokens, and burning $y$ tokens returns $R_t(y) = y \nav_t$ of collateral. With reserves $A_t$ and supply $S_t$, the coverage ratio is
\begin{equation}
\Gamma_t \;=\; A_t / (S_t \cdot \nav_t),
\label{eq:coverage}
\end{equation}
enforced above $\gamma_{\min} > 1$ on every state-changing call. \sysname is thus an index-linked redeemable claim against a fiat-backed reserve, with liabilities that move with the index; closer in spirit to an ETF redemption mechanism with a cost-driven NAV than to a fixed-peg stablecoin.

Four controls compose Algorithm~\ref{alg:vault}'s pre-call check. \emph{Coverage gating:} mint requires $\Gamma_t \ge \gamma_{\min}$ post-application; redeem checks coverage, but, when violated, queues rather than reverts so the claim is preserved. \emph{Adaptive mint throttle:} the per-epoch mint cap contracts with coverage headroom $h_t = \max(0, \Gamma_t - \gamma_{\min})$ via $C^{\text{mint}}_t = C^{\text{mint}}_0 \cdot \min(1, h_t / h^\star)$, so new issuance cannot eat the buffer that protects existing holders; the redeem cap $C^{\text{red}}_t = C^{\text{red}}_0$ stays constant so redemption capacity is preserved under stress. \emph{Auto-pause:} \textsc{StalePause} halts mint when the oracle is beyond $\tau$; \textsc{Paused} halts all state changes when $\Gamma_t < \gamma_{\text{pause}} \le \gamma_{\min}$. \emph{Defense-in-depth:} the on-chain drift cap duplicates the off-chain bound of equation~\ref{eq:smooth}. Together, these shift shock absorption from holders onto the would-be minter.

\begin{algorithm}[t]
\caption{\textsc{VaultOperation}: mint, redeem, and pre-call risk check.}
\label{alg:vault}
\begin{algorithmic}[1]
\Require op $\in$ \{\textsc{Mint}, \textsc{Redeem}\} with amount $z$; oracle $(\bar{I}_t, t_{\text{on}})$; vault $(A_t, S_t)$; risk parameters $\gamma_{\min}, \gamma_{\text{pause}}, \delta_{\max}, \tau$, base caps $(C^{\text{mint}}_0, C^{\text{red}}_0)$, headroom $h^\star$
\Statex \emph{// pre-call risk-control step}
\If{$\text{now} - t_{\text{on}} > \tau$} \textsc{StalePause} $\gets$ true \EndIf
\State $\nav_t \gets \bar{I}_t / I_0$; $\Gamma_t \gets A_t / (S_t \cdot \nav_t)$
\If{$\Gamma_t < \gamma_{\text{pause}}$} \textsc{Paused} $\gets$ true; emit \textsc{CoverageBreach} \EndIf
\State $h_t \gets \max(0, \Gamma_t - \gamma_{\min})$; $C^{\text{mint}}_t \gets C^{\text{mint}}_0 \cdot \min(1, h_t / h^\star)$; $C^{\text{red}}_t \gets C^{\text{red}}_0$; reset epoch usage; drain pending-redeem queue under coverage and $C^{\text{red}}_t$
\Statex \emph{// state-changing op}
\If{op $=$ \textsc{Mint}}
    \If{\textsc{Paused} or \textsc{StalePause}} \textbf{revert} \EndIf
    \State $\Delta S \gets z / \nav_t$
    \If{$U^{\text{mint}}_t + \Delta S > C^{\text{mint}}_t$} \textbf{revert} \Comment{rate limit} \EndIf
    \If{$(A_t + z) / ((S_t + \Delta S) \nav_t) < \gamma_{\min}$} \textbf{revert} \Comment{coverage} \EndIf
    \State commit $A_t \mathrel{+}= z$, $S_t \mathrel{+}= \Delta S$, $U^{\text{mint}}_t \mathrel{+}= \Delta S$; transfer $\Delta S$ \sysname; emit \textsc{Minted}
\Else \Comment{op = \textsc{Redeem}}
    \If{\textsc{Paused} or \textsc{StalePause}} enqueue $(z, \text{caller})$; \textbf{return} \EndIf
    \State $\Delta A \gets z \cdot \nav_t$
    \If{$U^{\text{red}}_t + \Delta A > C^{\text{red}}_t$ or $(S_t - z > 0$ and $(A_t - \Delta A) / ((S_t - z) \nav_t) < \gamma_{\min})$}
        \State enqueue $(z, \text{caller})$; \textbf{return} \Comment{soft rate limit and coverage check}
    \EndIf
    \State burn $z$ \sysname; transfer $\Delta A$; commit $A_t \mathrel{-}= \Delta A$, $S_t \mathrel{-}= z$, $U^{\text{red}}_t \mathrel{+}= \Delta A$; emit \textsc{Redeemed}
\EndIf
\end{algorithmic}
\end{algorithm}
\vspace{-0.3cm}

\subsection{Settlement Semantics}
\label{sec:design:settle}

\sysname is a standard transferable token, and this subsection is the entire payoff of tokenizing the cost anchor: what \emph{trading} the anchor actually looks like at the smart-contract layer. Agents quote services in \sysname and pay through ordinary transfers; because every agent reads the same NAV, quotes are directly comparable in compute units, eliminating the trusted middleware required by Baseline~A in Section~\ref{sec:model:why-token}.

The \emph{atomic multi-hop helper} handles delegation chains. Given a vector of (recipient, amount) tuples, it performs all transfers in a single transaction; any revert (insufficient balance, exceeded budget, or a downstream subtask failure signaled by a participating contract) reverts the entire bundle. This is a thin construction over standard transfer logic that is unavailable in Baseline~A, where each hop converts off-chain via a private index. An \emph{escrow} contract holds \sysname per task and releases funds on a signed receipt from the executing agent or refunds on timeout, again as one atomic transaction. A workflow originator can additionally pre-commit a maximum \sysname budget $B^\star$ to the escrow, which rejects any subtask payment that would exceed the remaining budget at the current NAV, namely a contract-enforced, compute-aligned budget constraint that internal accounting cannot supply.

\sysname coexists with fiat-backed stablecoins: the reserve is in USDC (or analogous collateral), and mint/redeem convert between USDC and \sysname at NAV. Hybrid agents hold USDC for outside-system payments and \sysname for compute-aligned coordination inside the agent economy. The resulting property set (shared NAV, atomic multi-hop settlement, contract-enforced budgets, auditable reserve, and risk state) is exactly what motivates moving the unit of account on-chain. A trusted off-chain service can replicate the first three properties under operator trust, but only an on-chain implementation provides the fourth and the censorship-resistance that follows. Section~\ref{sec:security} returns to this point and states what each layer guarantees against the adversary of Section~\ref{sec:model:threat}.

\section{Security Analysis}
\label{sec:security}

We analyze \sysname under the threat model of Section~\ref{sec:model:threat} and the design of Section~\ref{sec:design}. Section~\ref{sec:security:bounds} states bounded-adversary properties for the four security goals (G1) through (G4); Section~\ref{sec:security:scenarios} walks through the principal attack scenarios; Section~\ref{sec:security:limits} lists what the design does not cover. Bounds depend on parameters chosen by governance; Section~\ref{sec:eval:adversarial} estimates them empirically on the running prototype.

\subsection{Bounded-Adversary Properties}
\label{sec:security:bounds}

\textbf{Index integrity (G1).} Let $C^{\text{honest}}_{m,t}$ denote the honest reference cost for model $m$ at epoch $t$ (the representative cost that would emerge from prices honest vendors post) and let $\mathcal{A}$ control $f$ of $n_m$ vendors for model $m$. With the median estimator (equation~\ref{eq:med}) and $f < n_m / 2$, per-model error is bounded by the spread of \emph{honest} reports, independently of how extreme adversarial reports are.
\begin{lemma}[Median breakdown]
\label{lem:median}
If $f < n_m / 2$, then for every adversarial strategy
\[
\bigl|\tilde{C}^{\text{med}}_{m,t} - C^{\text{honest}}_{m,t}\bigr| \;\le\; \max_{v \in V^{\text{honest}}(m)} \bigl|C_{m,v,t} - C^{\text{honest}}_{m,t}\bigr|.
\]
\end{lemma}
\begin{proposition}[Index integrity]
\label{prop:index-integrity}
Let $\epsilon_m^{\text{hon}} = \max_{v \in V^{\text{honest}}(m)}\\ |C_{m,v,t} - C^{\text{honest}}_{m,t}|$ and assume $f < n_m / 2$ for every active $m$. The raw index satisfies $|I_t - I_t^{\text{honest}}| \le \sum_{m \in M^\star} w_m \epsilon_m^{\text{hon}}$.
\end{proposition}
The trimmed mean has breakdown $q$ with a corresponding bound on residual adversarial contribution; MAD-filtered mean retains $50\%$ breakdown and tightens tail sensitivity through $\kappa$, with an adversary conforming to the median/MAD envelope biasing the estimate by at most $\kappa \cdot \operatorname{MAD}$. The threshold $n_{\min}$ guarantees breakdown for every active model.

\textbf{NAV faithfulness (G2).} Let $I_t^\star$ be the index an honest oracle would publish and $\eta_t = \bar{I}_t - I_t^\star$ the publication deviation, bounded by whichever is smaller: the deviation any honest committee member would have produced, or the per-epoch drift cap $\delta_{\max}$.
\begin{proposition}[NAV deviation bound]
\label{prop:nav}
With smoothing $\lambda$, drift cap $\delta_{\max}$, and oracle staleness at most $\tau$, for any $\theta \in [t, t+\tau]$,
\[
\Bigl|\nav_\theta - \frac{I_\theta^\star}{I_0}\Bigr| \;\le\; \frac{1}{I_0}\Bigl(\sum_{s=0}^{\infty}(1-\lambda)\lambda^s |\eta_{t-s}| \;+\; \delta_{\max} \bar{I}_{t-1} \;+\; \mathrm{drift}(\tau)\Bigr),
\]
where $\mathrm{drift}(\tau)$ is the maximum honest movement of $I^\star$ over a window of length $\tau$.
\end{proposition}
The bound decomposes into attenuated past publication errors via the EMA, the per-epoch drift cap, and the staleness-window movement of the honest index. Tightening $\lambda, \delta_{\max}, \tau$ reduces NAV deviation at the cost of responsiveness or liveness: these are the knobs exposed to governance. Because the on-chain drift cap duplicates the off-chain bound, even a single-epoch majority compromise of the committee cannot exceed $\delta_{\max}$ in one step.

\textbf{Redeemability under stress (G3).} Let $g_t = \bar{I}_t / \bar{I}_{t-1} - 1$, so $|g_t| \le \delta_{\max}$. Let $\rho \ge 0$ be a reserve replenishment rate ($A_{t+1} \ge A_t(1+\rho)$ when mint and redeem are net zero). Algorithm~\ref{alg:vault} allows mint only while post-application coverage $\ge \gamma_{\min}$.
\begin{lemma}[Bounded coverage decay]
\label{lem:coverage}
Holding supply constant and ignoring redemption,
$\Gamma_{t+1} \ge \Gamma_t (1+\rho)/(1+\delta_{\max}).$
If $\rho \ge \delta_{\max}$ then coverage is non-decreasing.
\end{lemma}
\begin{theorem}[Solvency under bounded stress]
\label{thm:solvency}
With $\Gamma_0 \ge \gamma_{\min}$, per-epoch index growth bounded by $\delta_{\max}$, replenishment rate $\rho$, and the redemption queue processed under Algorithm~\ref{alg:vault}, $\Gamma_t \ge \Gamma_0 \bigl((1+\rho)/(1+\delta_{\max})\bigr)^{t}$. For any horizon $T$, $\Gamma_T \ge \gamma_{\min}$ provided $\rho \ge (1+\delta_{\max})(\gamma_{\min}/\Gamma_0)^{1/T} - 1$, which becomes $\rho \ge \delta_{\max}$ in the limit $\Gamma_0 \to \gamma_{\min}$.
\end{theorem}
A deployable \sysname must either start with substantial headroom $\Gamma_0 \gg \gamma_{\min}$ or adopt a yield-bearing reserve policy with $\rho \gtrsim \delta_{\max}$; this is the same balance-sheet constraint familiar from index-linked liabilities, and is why \sysname is presented as collateral-backed rather than algorithmically stabilized. A coordinated redemption rush is bounded by $C^{\text{red}}_t$ and the coverage check; queueing converts a rush into bounded FIFO outflows and avoids the bank-run dynamic of unbounded redemption while preserving the claim.

\textbf{Liveness and freshness (G4).} The oracle protocol admits any honest update within one confirmation and rejects updates older than $\tau$ through the staleness gate. Consumers read both $\bar{I}_t$ and $t_{\text{on}}$ and detect violations directly. When \textsc{StalePause} fires, mint reverts and redeem is queued: the protocol trades temporary mint capacity for refusal to act on untrusted state. Under base-chain censorship resistance, an honest quorum can always re-establish freshness.

\subsection{Adversarial Scenarios and Defenses}
\label{sec:security:scenarios}

\textbf{S1. Stale or delayed oracle data.} The adversary delays submissions. The staleness gate $\tau$ triggers \textsc{StalePause}, halting mint and queueing redeem; honest reporters recover by submitting a fresh attestation. During the pause, agents cannot mint: the protocol prefers refusal to action under an untrusted state.

\textbf{S2. Manipulated price inputs and Sybil vendors.} The adversary controls $f$ endpoints and posts biased prices. The median (Lemma~\ref{lem:median}) tolerates $f < n_m / 2$ with bounded error; $n_{\min}$ excludes models with insufficient honest sources; $h_t$ in $\sigma_t$ enables ex-post audit. Sybil \emph{addition} is restricted by the governance-controlled vendor registry. A coordinated posting that mimics honest dispersion can shift the median within the honest envelope; this residual is bounded by Proposition~\ref{prop:index-integrity}.

\textbf{S3. Strategic provider repricing.} A vendor with strong basket weight changes its public price to influence the index (analogous to benchmark-inclusion gaming). Smoothing and the drift cap attenuate single-vendor moves; governance can rebalance weights or remove the vendor under a timelock. Repeated small adjustments within the cap can still influence the long-run index; the design exposes this as a parameter rather than hiding it.

\textbf{S4. Mint/redeem MEV.} An adversary front-runs the oracle update transaction with a redeem (when the index is about to fall) or mint (when it is about to rise) and back-runs with the inverse. $\delta_{\max}$ bounds per-epoch sandwich profit; per-epoch mint and redeem caps bound any single attacker's turn; optional commit/reveal of oracle updates and per-block rate limits further reduce extractable value. Small per-epoch arbitrage is unavoidable in a fully transparent system; this is a bounded tax rather than an exploit.

\textbf{S5. Redemption rush under index spike.} A coordinated wave of redeem calls follows a permitted index increase. The coverage check and per-epoch redeem cap convert the rush into bounded outflows; FIFO queueing preserves the ordering; Theorem~\ref{thm:solvency} bounds when $\Gamma_t$ stays above $\gamma_{\min}$. If index growth persistently exceeds replenishment, queued redemptions extend; auto-pause and governance pause provide the explicit fallback.

\textbf{S6. Governance capture or proposal griefing.} Malicious basket or parameter changes, or a flooded proposal queue. The timelock delays application, allowing observation and counter-action; quorum and voting rules require the attacker to share above the safety threshold; the emergency pause is held under a multisig. Long-horizon governance attacks remain a known risk class; the design minimizes the governance surface rather than expanding it.

\textbf{S7. Hidden enterprise discounts (modeling gap).} Public list prices may overstate realized cost for some agents. The index is by construction over public prices; deviation is observable as a gap between the published index and realized expenditure, which agents can price into quotes. This residual reflects the protocol's choice to track a transparent but imperfect proxy.

\subsection{Out of Scope}
\label{sec:security:limits}

\sysname does not defend against: compromise of a vendor majority ($f \ge n/2$ defeats any breakdown bound, and no robust aggregator can recover a faithful price from majority-adversarial inputs); correlated regulatory shutdown of reserve custodians; bugs in the base chain or threshold-signature library; long-horizon coordinated governance capture above the quorum; or off-chain coercion of agents. These are system-level limitations (Section~\ref{sec:limitations}). The composition of (G1) through (G4) is what justifies the on-chain tokenized design over the internal-indexing baseline of Section~\ref{sec:model:why-token}: each property is enforced by code in a specific layer, with parameters that are visible and tunable by governance and with empirical bounds estimated in Section~\ref{sec:evaluation}.

\section{Prototype and Implementation}
\label{sec:prototype}

We instantiate the protocol of Section~\ref{sec:design} as a deployable prototype on a local Ethereum-compatible network, with an off-chain index calculator and a multi-agent client layer. The prototype is deliberately minimal: it implements the security-relevant logic (NAV-based mint/redeem, coverage check, oracle freshness, mint throttle, pause) and omits ancillary features such as cross-chain bridging, leveraged positions, or multi-collateral support.

\subsection{Smart Contract Stack}
\label{sec:proto:contracts}

The smart-contract stack comprises five contracts with disjoint responsibilities. \texttt{IndexOracle} stores $\bar{I}_t$, $t_{\text{on}}$, the active basket version, and $h_t$; it exposes a writer-authenticated \texttt{update} that rejects backdated, drift-violating, or unauthorized submissions, a \texttt{read} accessor, and \texttt{isStale($\tau$)}. \texttt{ClawCoinToken} is an ERC-20-compatible token whose \texttt{mint}/\texttt{burn} hooks are restricted to the vault; standard transfer, allowance, and DeFi composability work through the unmodified interface. \texttt{MintRedeemVault} implements Algorithm~\ref{alg:vault}: it holds reserve collateral, tracks per-epoch cap usage, maintains the pending-redeem queue, reads $\bar{I}_t$ from the oracle, and consults the risk manager for caps and pause flags. \texttt{RiskManager} stores $\gamma_{\min}, \gamma_{\text{pause}}, \delta_{\max}, \tau$, base caps, and $h^\star$, holds the pause flags, and is invoked as the precondition check. \texttt{Governance} manages the basket, workload weights, oracle membership, and risk parameters under a timelock with explicit minimum delay, and holds the emergency pause under a separate multisig. The chain trusts writer authentication, coverage, and rate-limit enforcement, and timelock invariants; it does not trust the correctness of $\bar{I}_t$ with respect to off-chain markets (auditable via $h_t$) or the absence of MEV in the surrounding chain (handled by per-epoch caps, drift cap, and optional commit/reveal updates).

\subsection{Off-Chain Index Calculator}
\label{sec:proto:offchain}

Each oracle node runs Algorithm~\ref{alg:publication}. Versioned per-vendor adapters normalize public price schedules into $(P^{\text{in}}_{m,v,t}, P^{\text{out}}_{m,v,t})$; prices are fetched in parallel with timeouts, retries, and freshness timestamps; per-pair cost follows equation~\ref{eq:cost}; robust per-model aggregation (equations~\ref{eq:med} through \ref{eq:mad}) is gated by $n_{\min}$; the basket, EMA, and drift-capped publication follow equation~\ref{eq:smooth}. Before submission, the node builds a Merkle tree over observations, signs $\sigma_t$, and dispatches it to the on-chain oracle (committee path) or the aggregator (DON path). Vendor adapters follow the same timelocked governance path as basket changes; periodic replay jobs reconstruct the index from stored roots for third-party audit.

\subsection{Deployment Target}
\label{sec:proto:chain}

We target an Ethereum L2. Multi-hop delegation requires many small-value sub-second transfers, for which L2 per-transaction costs make the atomic settlement primitive practical; ERC-20 and standard signature verification keep the prototype surface area on widely audited libraries; and L2s host the same DeFi primitives (escrow, AMMs, lending) that hybrid \sysname/USDC agents may use. The protocol is otherwise chain-agnostic: the security analysis does not depend on a specific execution environment beyond a censorship-resistant base chain.

\subsection{OpenClaw Integration}
\label{sec:proto:openclaw}

We integrate \sysname with OpenClaw, a multi-agent system in which collaborating agents quote, execute, and delegate compute-intensive subtasks over a shared messaging bus. OpenClaw is described and evaluated separately; here we state only the integration interface. Each agent is provisioned with an EOA and a \sysname balance; its protocol-level identity is its address, and signed messages bind off-chain quotes and receipts to on-chain settlement.

OpenClaw exposes a pluggable settlement interface with three backends. \texttt{usdc-fiat} quotes and pays in USDC with fiat denominated budgets (the Fiat baseline). \texttt{usdc-internal-index} quotes and pays in USDC but lets each agent maintain a private compute cost index and convert at quote time (the USDC + internal-indexing baseline of Section~\ref{sec:model:why-token}). \texttt{clawcoin} quotes, pays, and settles in \sysname, using the atomic multi-hop helper for delegation chains and the escrow contract for asynchronous execution. Because the same workflow, policies, and task graph are replayed under each backend with no change to agent logic, the regime comparison is clean. Under \texttt{clawcoin}, an orchestrator gathers \sysname-denominated quotes, selects executors subject to a pre-committed \sysname budget, locks the budget in escrow, dispatches subtasks, and on receipt of signed completions releases payment via the atomic bundle, with the entire chain committing or reverting together.

\section{Evaluation}
\label{sec:evaluation}

\sysname is presented as a compute-cost-aligned unit of account for decentralized agent economies, and the evaluation is organized around that claim. Section~\ref{sec:eval:setup} describes the prototype, simulator, and OpenClaw testbed. Section~\ref{sec:eval:single} reports single-agent stability, pricing, and market-feasibility properties under four monetary regimes. Section~\ref{sec:eval:workflow} is the main result: multi-agent workflows, long-horizon economy survival, and external resource procurement on the OpenClaw testbed, where the value of moving the unit of account on-chain is most visible. Section~\ref{sec:eval:adversarial} compresses the adversarial and solvency tests into three minimal sanity checks confirming that the risk module behaves as specified.

\subsection{Setup and OpenClaw Testbed}
\label{sec:eval:setup}

\textbf{On-chain prototype.} The five-contract stack of Section~\ref{sec:proto:contracts} runs on a local Ethereum-compatible development network. The off-chain calculator of Section~\ref{sec:proto:offchain} runs against synthetic vendor adapters whose price trajectories are controllable from the simulator, so the same trajectory replays across regimes.

\textbf{Multi-agent simulator.} A market simulator drives task arrivals, quoting, execution, and settlement under a chosen regime; provider costs evolve through stochastic fluctuations and discrete repricing shocks. Each agent maintains a treasury, provider mix, markup policy, and quoting rule, and may participate in delegation chains.

\textbf{OpenClaw testbed.} For workflow-level experiments we deploy OpenClaw on a dedicated hardware testbed. Each OpenClaw agent runs in its own Docker container on a single x86 host; each container holds exactly one agent with its wallet, settlement adapter, role-specific prompt and tooling, and a local SQLite store. Inter-agent communication uses a shared message bus. The host has no GPU: inference is provided by a separate server with four NVIDIA RTX 6000 Pro GPUs (96\, GB VRAM) running an OpenAI-compatible endpoint that serves GLM-4.7 Flash. All agents issue completions through this single endpoint, so the cognitive layer is held constant across regimes and only the settlement backend changes between runs.

\textbf{OpenClaw agent roles.} We instantiate six roles: a \emph{planner} decomposes tasks, quotes, and orchestrates delegation; a \emph{retriever} performs structured search over local/remote indices; a \emph{tool-use} agent invokes external APIs and sandboxes; a \emph{coder} runs short programs against a Python sandbox; a \emph{verifier} grades artifacts against acceptance criteria; and a \emph{synthesizer} merges retrieved evidence and tool outputs into the final response. Each role binds a fixed prompt, tool whitelist, context window, and published per-call \sysname price quote signed along with completion receipts. Roles are otherwise homogeneous: same GLM-4.7 Flash backend, same wallet/identity primitives, same adapter interface.

\textbf{Settlement backends and regimes.} The OpenClaw runtime exposes \texttt{usdc-fiat}, \texttt{usdc-internal-index}, and \texttt{clawcoin} (Section~\ref{sec:proto:openclaw}). For simulator experiments we compare four regimes: Fiat (USDC), Raw-cost (quotes track instantaneous provider-specific prices without a shared numeraire), USDC + internal indexing (per-agent private index, Baseline~A of Section~\ref{sec:model:why-token}), and \sysname. Metrics group into \emph{stability} (capacity variance, drawdown, recovery, CoV), \emph{coordination} (quote volatility, repricing frequency, cross-agent dispersion, price consistency), \emph{market/workflow feasibility} (acceptance/completion, budget-overrun, partial-settlement, latency), and \emph{risk checks} (oracle-staleness, vendor-manipulation, redemption-burst).

\subsection{Single-Agent Stability, Pricing, and Market Feasibility}
\label{sec:eval:single}

We first verify in the simulator that \sysname stabilizes per-agent budgets, reduces pricing friction across heterogeneous agents, and improves market feasibility under explicit budget constraints. These are properties that internal indexing can partially recover per-agent; the multi-agent results of Section~\ref{sec:eval:workflow} then expose what internal indexing cannot do.

\begin{figure}[t]
\centering
\includegraphics[width=0.85\linewidth]{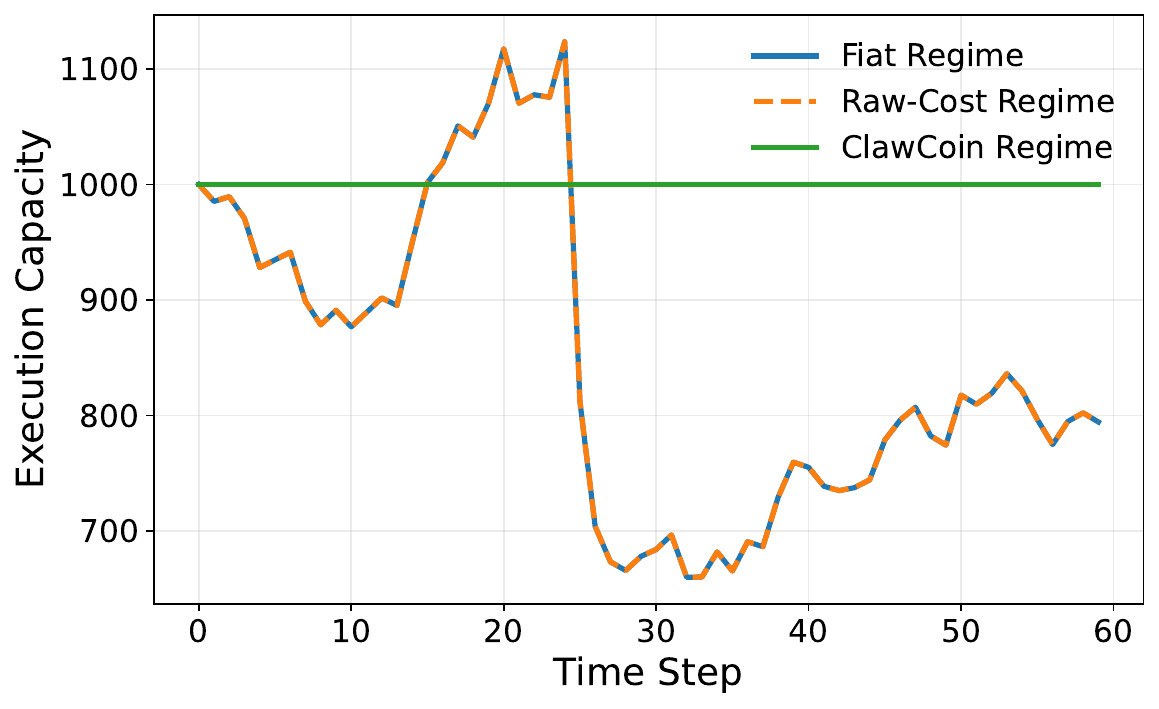}
\Description{Line plot of execution capacity over time for each monetary regime, showing that \sysname preserves near-constant capacity under cost shocks while fiat and raw-cost accounting exhibit large drops and slow recovery.}
\caption{Execution capacity of a fixed nominal treasury under dynamic compute costs. Fiat and raw-cost accounting suffer substantial purchasing-power instability after shocks; \sysname preserves stable capacity through index-linked denomination.}
\label{fig:budget_stability}
\end{figure}

\textbf{Execution capacity.} A representative agent is endowed with a fixed nominal treasury under each regime, and at each step, we measure the standardized computational work the treasury can fund as the underlying compute index $\bar{I}_t$ moves through stochastic fluctuations and discrete shocks. Figure~\ref{fig:budget_stability} and Table~\ref{tab:eval:capacity} tell a single qualitative story: \sysname holds machine purchasing power essentially constant by construction; USDC + internal indexing recovers most of the per-agent stability because the agent privately reprices against its own basket; Fiat and Raw-cost both suffer large drawdowns and slow recovery because their nominal balances do not move with the cost they must pay. The gap between internal indexing and \sysname is small in the single-agent view and widens sharply once coordination enters.

\begin{table}[t]
\caption{Single-agent execution capacity under compute-cost shocks. Capacity normalized so that the noise-free reference treasury equals 1000.}
\label{tab:eval:capacity}
\small
\centering
\begin{tabular}{@{}l r r r r r@{}}
\toprule
Regime & Mean & Var. & Drawdown & Recov. & CoV \\
\midrule
Fiat            & 918  & 12{,}840 & 24.7\% & 19 & 0.124 \\
Raw-cost        & 884  & 17{,}310 & 31.5\% & 24 & 0.149 \\
USDC + internal & 987  & 1{,}840  & 6.8\%  & 6  & 0.043 \\
\sysname        & 1000 & 320      & 1.9\%  & 1  & 0.018 \\
\bottomrule
\end{tabular}
\vspace{-0.3cm}
\end{table}

\begin{figure}[t]
\centering
\includegraphics[width=0.85\linewidth]{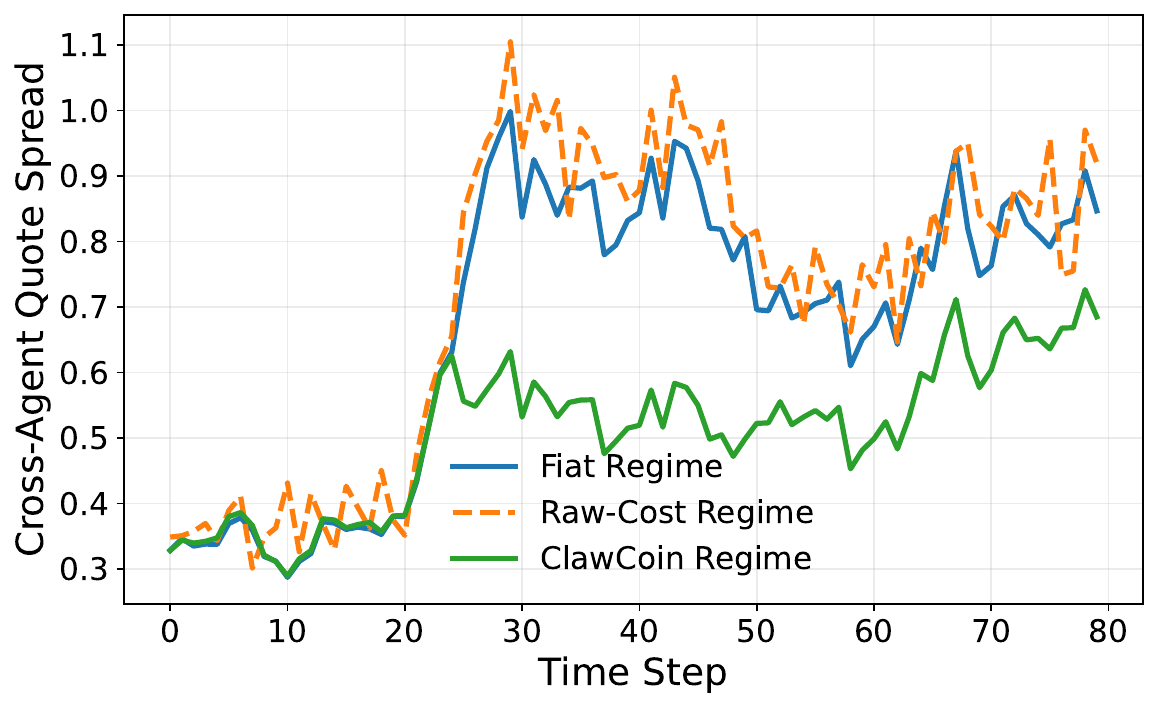}
\Description{Plot of cross-agent quote dispersion across regimes, showing raw-cost pricing with the widest spread and \sysname collapsing the spread by pricing quotes in a shared compute-aligned unit.}
\caption{Cross-agent quote dispersion across regimes. Raw-cost pricing exhibits the largest dispersion; \sysname reduces spread by internalizing common compute movements into the shared unit.}
\label{fig:pricing_spread}
\end{figure}

\begin{figure}[t]
\centering
\includegraphics[width=0.8\linewidth]{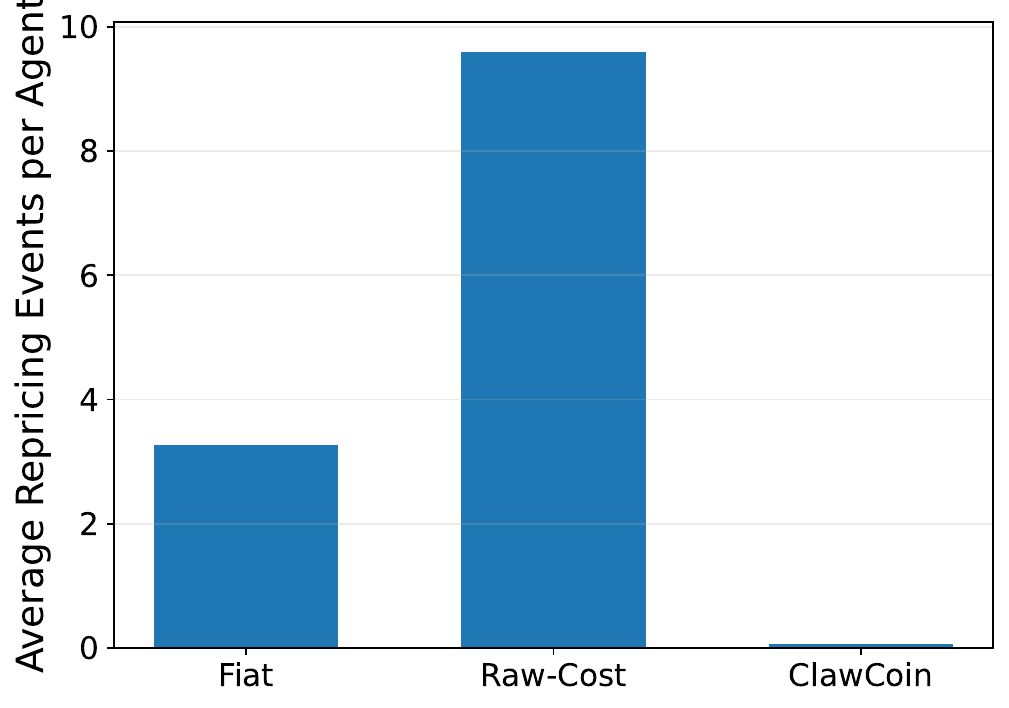}
\Description{Bar chart of the number of repricing events per agent in each monetary regime, with fiat and raw-cost producing the most repricings and \sysname producing the fewest.}
\caption{Repricing events per agent across regimes. Fiat and raw-cost require repeated nominal updates; \sysname absorbs common movements at the unit-of-account layer.}
\label{fig:repricing_frequency}
\end{figure}

\textbf{Pricing stability and repricing overhead.} A market of heterogeneous service-providing agents quotes a standardized request under common cost shocks; agents differ in provider mix and markup policy. Figures~\ref{fig:pricing_spread} through \ref{fig:repricing_frequency} visualize cross-agent dispersion and repricing behavior; Table~\ref{tab:eval:pricing} summarizes the four-regime comparison on five metrics. \sysname yields the lowest quote volatility, the fewest repricing events, and the tightest cross-agent dispersion. USDC + internal indexing improves over Fiat on volatility and repricing, but its cross-agent dispersion and price-consistency score stay materially worse than \sysname's because \emph{each agent uses its own basket}: absent a shared numeraire, per-agent stabilization cannot collapse quotes onto a common scale, even when each agent's own accounting is stable.

\begin{table}[t]
\caption{Pricing stability and repricing overhead.}
\label{tab:eval:pricing}
\footnotesize
\centering
\setlength{\tabcolsep}{3pt}
\begin{tabularx}{\columnwidth}{@{}l*{5}{>{\raggedleft\arraybackslash}X}@{}}
\toprule
Regime & Vol. & Repr./100 & Disp. & Consist. & Drift \\
\midrule
Fiat            & 0.118 & 26.4 & 0.141 & 0.77 & 0.084 \\
Raw-cost        & 0.153 & 31.9 & 0.187 & 0.69 & 0.109 \\
USDC + int.     & 0.074 & 14.2 & 0.093 & 0.86 & 0.049 \\
\sysname        & 0.039 & 6.8  & 0.041 & 0.94 & 0.021 \\
\bottomrule
\end{tabularx}
\vspace{-0.3cm}
\end{table}

\textbf{Task-market feasibility under budget constraints.} Customers submit tasks with heterogeneous willingness to pay; each task may require one or more service stages from different agents; a task is accepted if the aggregate quote fits within the customer's budget. Table~\ref{tab:eval:market} shows that \sysname raises acceptance and completion and collapses budget overruns, primarily because compute-aligned quotes match compute-aligned customer budgets. USDC + internal indexing narrows the gap to Fiat but cannot eliminate budget mismatches: customers and providers still reason in different units.

\begin{table}[t]
\caption{Task-market feasibility. Mean price normalized to the Fiat regime.}
\label{tab:eval:market}
\footnotesize
\centering
\setlength{\tabcolsep}{3pt}
\begin{tabularx}{\columnwidth}{@{}l*{5}{>{\raggedleft\arraybackslash}X}@{}}
\toprule
Regime & Accept & Compl. & Overrun & Price & Rej. \\
\midrule
Fiat            & 78.6\% & 72.4\% & 6.9\% & 1.00x & 14.8\% \\
Raw-cost        & 73.1\% & 67.2\% & 9.8\% & 0.98x & 18.6\% \\
USDC + int.     & 83.5\% & 79.9\% & 4.1\% & 1.03x & 10.2\% \\
\sysname        & 89.8\% & 87.1\% & 0.8\% & 1.02x & 5.4\% \\
\bottomrule
\end{tabularx}
\end{table}

\subsection{Multi-Agent Workflows and OpenClaw Integration}
\label{sec:eval:workflow}

The multi-agent results are the main system contribution of the evaluation. They directly test the claim of Section~\ref{sec:model:why-token} that internal indexing alone cannot supply atomic multi-hop settlement, contract-enforced budgets, and a shared numeraire for cross-agent coordination.

\textbf{Workflow-level evaluation.} The simulator is extended with explicit multi-step delegation chains of depth $D \in \{1,2,4,8\}$. Tasks have stage-level subtasks with stage-level prices, budgets are pre-committed at submission, and mid-flight cost shocks may push aggregate cost above the budget. Under \sysname, the workflow uses the atomic multi-hop helper and the escrow budget gate (Section~\ref{sec:design:settle}); under USDC + internal indexing, each hop converts via the originator's private index with off-chain reconciliation; under Fiat and Raw-cost each hop denominates in USDC. Table~\ref{tab:eval:workflow} reports the aggregate comparison, and Table~\ref{tab:eval:workflow-depth} reports the depth breakdown. Two qualitative findings stand out. First, \sysname drives partial-settlement to zero by construction (any failed hop reverts the bundle) and eliminates budget overruns through the pre-commit gate; both are contract-level properties that the non-tokenized baselines cannot replicate without trusted middleware. Second, failure rates compound with depth in every regime, but the slope differs sharply: Fiat and Raw-cost grow steeply because mid-flight shocks compound across hops with no atomic primitive to bind them; USDC + internal indexing grows more slowly but remains vulnerable to coordination failures at $D=8$; \sysname grows almost linearly with depth because each hop settles in the same compute-aligned unit and the bundle is all-or-nothing.

\begin{table}[t]
\caption{Workflow-level evaluation across delegation depths $D \in \{1,2,4,8\}$. Settlement latency normalized to Fiat.}
\label{tab:eval:workflow}
\scriptsize
\centering
\setlength{\tabcolsep}{2pt}
\begin{tabularx}{\columnwidth}{@{}l*{6}{>{\raggedleft\arraybackslash}X}@{}}
\toprule
Regime & Fail & Over. & Partial & Err. & Lat. & p95 \\
\midrule
Fiat            & 9.8\%  & 7.1\%  & 6.4\% & 0.143 & 1.00x & 1.42x \\
Raw-cost        & 14.6\% & 10.9\% & 9.7\% & 0.188 & 0.96x & 1.51x \\
USDC + int.     & 6.2\%  & 4.8\%  & 5.9\% & 0.091 & 1.12x & 1.39x \\
\sysname        & 2.1\%  & 0.0\%  & 0.0\% & 0.037 & 1.08x & 1.17x \\
\bottomrule
\end{tabularx}
\vspace{-0.3cm}
\end{table}

\begin{table}[t]
\caption{Workflow failure rate by delegation depth.}
\label{tab:eval:workflow-depth}
\small
\centering
\begin{tabular}{@{}l r r r r@{}}
\toprule
Depth & Fiat & Raw & USDC + int. & \sysname \\
\midrule
$D=1$ & 3.2\%  & 5.8\%  & 2.7\%  & 1.1\% \\
$D=2$ & 6.7\%  & 10.9\% & 4.4\%  & 1.8\% \\
$D=4$ & 11.8\% & 16.5\% & 7.0\%  & 2.5\% \\
$D=8$ & 17.6\% & 24.8\% & 10.8\% & 3.2\% \\
\bottomrule
\end{tabular}
\vspace{-0.3cm}
\end{table}

\textbf{End-to-end OpenClaw workflows.} We replay representative workflows on the OpenClaw testbed under each backend without changes to agent logic. Three workflow categories are used: document analysis (planner $\to$ retriever $\to$ synthesizer), structured retrieval (planner $\to$ retriever $\to$ verifier $\to$ synthesizer), and multi-tool research (planner $\to$ retriever $\to$ tool-use $\to$ coder $\to$ verifier $\to$ synthesizer). Table~\ref{tab:eval:openclaw-e2e} reports workflow success, protocol revert rate, the mean number of distinct agents per task, and realized cost drift against the originator's pre-commit estimate. \texttt{clawcoin} improves workflow success and collapses protocol reverts and cost drift. The gain comes entirely from the settlement and budget layer, not from any change in the cognitive layer: all three backends share the same GLM-4.7 Flash endpoint and the same per-role prompts, so what shrinks are the mid-flight repricing mismatches and reconciliation failures that the non-atomic baselines expose.

\begin{table}[t]
\caption{End-to-end OpenClaw workflow evaluation across settlement backends.}
\label{tab:eval:openclaw-e2e}
\footnotesize
\centering
\setlength{\tabcolsep}{3pt}
\begin{tabularx}{\columnwidth}{@{}l*{4}{>{\raggedleft\arraybackslash}X}@{}}
\toprule
Backend & Success & Revert & Agents & Drift \\
\midrule
\texttt{usdc-fiat}            & 81.5\% & 8.4\% & 3.6 & 0.116 \\
\texttt{usdc-internal-index}  & 84.8\% & 6.1\% & 3.7 & 0.081 \\
\texttt{clawcoin}             & 92.7\% & 2.3\% & 3.8 & 0.034 \\
\bottomrule
\end{tabularx}
\vspace{-0.3cm}
\end{table}

\textbf{Long-horizon economy survival.} To probe coordination value over longer horizons, we instantiate eight OpenClaw agents with heterogeneous initial treasuries and role specializations and run a 500-round market: tasks arrive stochastically, agents quote, the planner delegates, executors deliver, and treasuries update by equation~\ref{eq:treasury}. An agent is declared bankrupt and exits if its treasury cannot cover expected operating costs over the next three rounds. Table~\ref{tab:eval:openclaw-survival} shows that under \texttt{clawcoin}, more agents survive, median treasuries \emph{grow} rather than shrink, wealth distribution is less concentrated, and inter-agent trade volume and delegation rate rise substantially. The mechanism is not an external subsidy: every backend runs against the same task-arrival process and the same model. The mechanism is that a shared compute-aligned numeraire makes cross-agent quotes legible enough for delegation to be profitable, which raises equilibrium trade volume and avoids the cascading bankruptcies that follow from mispriced delegation under fiat denomination.

\begin{table}[t]
\caption{OpenClaw economy survival over 500 rounds. Trade volume normalized to the Fiat backend.}
\label{tab:eval:openclaw-survival}
\footnotesize
\centering
\setlength{\tabcolsep}{2pt}
\begin{tabularx}{\columnwidth}{@{}l*{5}{>{\raggedleft\arraybackslash}X}@{}}
\toprule
Backend & Surv. & $\Delta T$ & Gini & Vol. & Deleg. \\
\midrule
\texttt{usdc-fiat}            & 4/8 & $-18.7\%$ & 0.49 & 1.00x & 0.38 \\
\texttt{usdc-internal-index}  & 5/8 & $-9.4\%$  & 0.43 & 1.12x & 0.44 \\
\texttt{clawcoin}             & 6/8 & $+6.8\%$  & 0.38 & 1.34x & 0.57 \\
\bottomrule
\end{tabularx}
\vspace{-0.3cm}
\end{table}

\textbf{OpenClaw resource procurement.} We extend OpenClaw with a remote-execution adapter that exposes burstily priced GPU time as an executable subtask; three locally provisioned OpenClaw agents and three remotely provisioned OpenClaw agents run side by side, and the planner chooses between local and remote execution at each subtask. Table~\ref{tab:eval:openclaw-procurement} shows that the dominant pain point in the integration is not payment transport (USDC moves cleanly across backends) but \emph{procurement under a shared compute-aligned budget}. Under \texttt{clawcoin}, mid-task repricing incidents collapse, completed workflows rise, and spend predictability across identical workflow classes approaches one. The compute-aligned unit lets the planner reason about local-versus-remote substitution in a single budget unit, which is the missing-primitive argument of Section~\ref{sec:model:why-token} applied to OpenClaw's external resource market.

\begin{table}[t]
\caption{OpenClaw GPU-resource procurement across settlement backends.}
\label{tab:eval:openclaw-procurement}
\footnotesize
\centering
\setlength{\tabcolsep}{2pt}
\begin{tabularx}{\columnwidth}{@{}l*{5}{>{\raggedleft\arraybackslash}X}@{}}
\toprule
Backend & Attach & Deleg. & Repr. & Compl. & Pred. \\
\midrule
\texttt{usdc-fiat}            & 86.1\% & 79.4\% & 17.8\% & 76.8\% & 0.74 \\
\texttt{usdc-internal-index}  & 88.7\% & 82.5\% & 12.3\% & 80.6\% & 0.82 \\
\texttt{clawcoin}             & 95.4\% & 91.8\% & 4.1\%  & 89.9\% & 0.93 \\
\bottomrule
\end{tabularx}
\vspace{-0.3cm}
\end{table}

Across all four multi-agent experiments, the qualitative pattern is consistent: moving the unit of account on-chain is what converts compute-aligned \emph{accounting} into compute-aligned \emph{coordination}. USDC + internal indexing recovers most of the per-agent stability but cannot collapse cross-agent dispersion, cannot atomically bind multi-hop delegation, cannot contract-enforce a shared compute budget, and cannot sustain cooperative market dynamics over long horizons. \sysname does all four, and the gap widens monotonically with delegation depth, agent heterogeneity, and time horizon.

\subsection{Risk Sanity Checks}
\label{sec:eval:adversarial}

The risk module is engineering hygiene, not the paper's main contribution. We therefore reduce the adversarial and solvency evaluation to three sanity checks that confirm the properties of Section~\ref{sec:security} hold on the running prototype, and defer larger adversarial sweeps to future work. Table~\ref{tab:eval:sanity} groups the three checks.

\textbf{Stale oracle.} We freeze the update channel. Beyond the staleness threshold, \texttt{IndexOracle.isStale} returns true, mint is rejected, and incoming redeem requests are queued rather than executed against an untrusted price. The contract resumes normal operation on the next fresh attestation, exactly as specified in Section~\ref{sec:design:vault}.

\textbf{Single-vendor manipulation.} A single vendor in a basket of five posts has a $+50\%$ biased input price. Under the mean estimator, the published index absorbs the bias proportionally; under the median (the protocol default), the deviation collapses by more than an order of magnitude, consistent with Lemma~\ref{lem:median}; with the on-chain drift cap engaged, per-epoch deviation is further bounded to the cap. The check is deliberately small: its role is to verify that the median-plus-drift-cap composition behaves as expected on the prototype.

\textbf{Redemption burst.} We drive the prototype through three index trajectories with a coordinated redemption rush: a mild shock with one drift-capped step, a two-step spike, and sustained capped growth combined with a low replenishment rate. Under the mild shock, the queue drains cleanly. Under the two-step spike, most redemptions are honored within a short horizon, and the rest queue. Under sustained growth with low replenishment, the auto-pause condition engages as predicted by Lemma~\ref{lem:coverage}. Across all three, no involuntary loss to redeemers is observed; the queue absorbs short rushes, and auto-pause is the last line of defense under sustained adverse motion.

\begin{table*}[t]
\caption{Risk sanity checks on the running prototype. All three confirm that the risk module behaves as designed.}
\label{tab:eval:sanity}
\small
\begin{tabularx}{\textwidth}{p{3.0cm} X X X X}
\toprule
Check & Condition & Observed behavior & Bound from \S\ref{sec:security} & Honored / safe \\
\midrule
Stale oracle    & Beyond staleness threshold (4 epochs)        & Mint rejected; redeem queued; no stale-state settlement & G4, Algorithm~\ref{alg:vault}            & 100\% queued; recovers next epoch \\
Vendor bias     & One vendor of five posts $+50\%$ input bias  & Mean: $+11.8\%$; median: $+1.9\%$; +drift cap: $+1.0\%$/epoch  & Lemma~\ref{lem:median}, Eq.~\ref{eq:smooth} & Median + cap suppress bias \\
Redemption burst & Mild / two-step spike / sustained capped growth & Queue peak 18 / 63 / 141; pause on the third         & Lemma~\ref{lem:coverage}, Theorem~\ref{thm:solvency} & 100\% / 92\% / 61\% within 5 epochs \\
\bottomrule
\end{tabularx}
\vspace{-0.3cm}
\end{table*}

These results are protocol hygiene. The system claim of \sysname rests on the multi-agent results of Section~\ref{sec:eval:workflow}, where moving the unit of account on-chain is what converts compute-aligned accounting into compute-aligned coordination.

\section{Discussion, Limitations, and Future Work}
\label{sec:limitations}

\subsection{Discussion}

\textbf{Operational asset, not speculative instrument.} \sysname is intended as operational infrastructure for decentralized agent economies, not as a retail investment product. Demand arises from transactional utility (agents need a stable unit for budgeting, quoting, and atomic multi-hop settlement of compute-bound workflows) rather than from appreciation expectations. Any transferable asset can be traded, but speculation is not required for the system to produce utility, and the risk module's per-epoch caps and pause logic deliberately bound secondary-market velocity.

\textbf{Cost-only by design.} As argued in Section~\ref{sec:design:index}, the index tracks inference cost rather than model quality, which is priced separately in markets through reputation, contracting, and observed outcomes. A consequence is that the unit of account alone cannot resolve information asymmetry about service quality: two agents that quote identical \sysname prices for the same workload may still deliver meaningfully different outputs, and the economy must layer reputational and contractual signals on top of the compute-aligned unit.

\textbf{Regulatory posture.} Because \sysname is redeemable against reserve collateral and may circulate as a transferable digital asset, its legal treatment will depend on jurisdiction and deployment model; it may fall under stored-value, prepaid-instrument, or reserve-backed-claim regimes in different settings, and consortium or enterprise deployments among known counterparties have a different profile from public circulation. We treat compliance architecture as a deployment-layer concern rather than a property of the monetary design itself.

\textbf{Toward fully machine-native pricing.} The current index ingests prices quoted in fiat, reflecting how most commercial inference is priced today and enabling transparent list-price construction. As compute markets evolve toward direct machine-to-machine pricing, native compute auctions, or tokenized GPU markets, the same pipeline can ingest those inputs without protocol redesign, and the unit becomes more fully machine-native.

\subsection{Limitations}

The index is defined over public list prices and cannot capture hidden enterprise discounts (Section~\ref{sec:security:scenarios}, S7), which bounds how closely the index tracks any individual agent's realized cost. The integrity bound in Proposition~\ref{prop:index-integrity} requires $f < n_m / 2$ honest vendors per active model, and the NAV bound in Proposition~\ref{prop:nav} assumes an honest oracle threshold; beyond these, guarantees degrade gracefully (auto-pause on detected drift; redemption queueing) but cannot be tightened by parameter choice alone. Theorem~\ref{thm:solvency} bounds coverage under bounded per-epoch growth and a positive replenishment rate, but persistent cost increases above the replenishment rate cannot be absorbed indefinitely by a passive collateral pool, so deployable systems require either substantial overcollateralization, sponsor-seeded buffer capital, or yield-bearing reserve management. The standardized workload classes approximate rather than replace real workloads, so agents heavily reliant on non-token components (caching, batching, hardware tiers, network tail latency) experience a weaker match; the basket can be widened but full alignment is unattainable in a publicly verifiable way. Finally, Section~\ref{sec:security:limits} enumerates the threats not covered (full vendor collusion, correlated regulatory shutdown, base-chain bugs, long-horizon governance capture) and the current evaluation uses a modest-scale prototype, simulator, and testbed; larger live deployments are left to future work.

\subsection{Future Work}

Several directions follow naturally. Zero-knowledge attestations binding index inputs to authenticated provider responses would upgrade Proposition~\ref{prop:index-integrity} to a bound rooted in cryptographic assumptions rather than vendor independence. Replacing the threshold committee with a permissionless DON (Pyth or UMA style) and quantifying the resulting latency/freshness/cost trade-offs is a natural deployment-track extension. Extending the workload basket to capture latency tiers, throughput SLAs, context-window pricing, and bundled enterprise contracts, and connecting it to tokenized GPU and inference auctions, would tighten the gap between the index and realized cost. On the analysis side, equilibrium analysis of provider strategic repricing, governance-attack incentives, and MEV around oracle updates would complement the bounded-adversary results of Section~\ref{sec:security}. An end-to-end OpenClaw deployment that closes the loop between Section~\ref{sec:eval:workflow} and a small live deployment, including regulated and consortium settings, is the most concrete operational next step.

\section{Conclusion}

Decentralized agent economies have an emerging payment substrate but lack a representation of machine-native value: a tradable, composable, contract-readable unit aligned with the computational cost that bottlenecks agent execution. We argued that internal cost indexing over stablecoin rails cannot supply this primitive in a decentralized setting, and we presented \sysname, a collateral-backed, compute-indexed tokenized unit of account whose four protocol layers (robust basket index, attested oracle, NAV-based mint/redeem vault with risk controls, and on-chain atomic multi-hop settlement) are designed under a concrete adversary model. We stated bounded-adversary properties for index integrity, NAV faithfulness, redeemability under stress, and liveness, and walked through the principal attack scenarios. A prototype on an Ethereum-compatible network, a multi-agent simulator, and a six-role OpenClaw testbed show that moving the unit of account on-chain is what converts compute-aligned accounting into compute-aligned \emph{coordination}: cross-agent quote dispersion collapses, multi-hop delegation settles atomically, contract-level budgets bind end-to-end, and long-horizon market dynamics sustain cooperative trade. \sysname is a security-engineered, on-chain, compute-cost-indexed unit of account, and a coherent, implementable step toward the missing representation primitive of the decentralized AI agent economy.

\end{document}